\documentclass[preprint,showpacs,preprintnumbers,amsmath,amssymb]{revtex4}

\usepackage{graphicx}
\usepackage{epsfig}		
\usepackage{dcolumn}
\usepackage{bm}
\usepackage{color}

\def\0{\mbox{\tiny $0$}}
\def\1{\mbox{\tiny $1$}}
\def\2{\mbox{\tiny $2$}}
\def\3{\mbox{\tiny $3$}}
\def\4{\mbox{\tiny $4$}}
\def\5{\mbox{\tiny $5$}}
\def\6{\mbox{\tiny $6$}}
\def\7{\mbox{\tiny $7$}}
\def\8{\mbox{\tiny $8$}}
\def\9{\mbox{\tiny $9$}}

\def\f14{\mbox{\tiny $\frac{1}{4}$}}
\def\infm{\mbox{\tiny $-\infty$}}
\def\infp{\mbox{\tiny $+\infty$}}

\def\ii{\mbox{\tiny $i$}}
\def\z{\mbox{\tiny $z$}}

\def\j{\mbox{\tiny $j$}}

\def\mi{\mbox{\tiny $-$}}

\begin{document}

\title{Influence of second-order corrections to the energy-dependence of neutrino flavor conversion formulae}

\author{A. E. Bernardini}
\email{alexeb@ifi.unicamp.br}
\author{M. M. Guzzo}
\email{guzzo@ifi.unicamp.br}
\affiliation{Instituto de F\'{\i}sica Gleb Wataghin, UNICAMP,\\
PO Box 6165, 13083-970, Campinas, SP, Brasil.}

\date{\today}

\begin{abstract}
We discuss the {\em intermediate} wave-packet formalism for analytically quantifying the energy dependence of the two-flavor conversion formula that is usually considered for analyzing neutrino oscillations and adjusting the focusing horn, target position and/or detector location of some flavor conversion experiments.
Following a sequence of analytical approximations where we consider the second-order corrections in a power series expansion of the energy, we point out a {\em residual} time-dependent phase which, in addition to some well known wave-packet effects, can subtly modify the oscillation parameters and limits.
In the present precision era of neutrino oscillation experiments where higher precision measurements are required, we quantify some small corrections in neutrino flavor conversion formulae which lead to a modified energy-dependence for $\nu_{\mu}\leftrightarrow\nu_{e}$ oscillations.
\end{abstract}

\pacs{02.30.Mv, 03.65.Pm, 14.60.Pq}
\keywords{Wave Packets - Flavor Oscillation - Neutrino}
\maketitle

Although neutrino physics has been fueled by the recent growth in quality and quantity of experimental data \cite{sol1,sol2,atm1,atm2,atm3,Boe01,Egu031,Egu032}, there are still open questions on the theoretical front \cite{Zub98,Alb03,Vog04,Beu03} which, in some cases,
concern with obtaining more refined parameters from flavor conversion formulae \cite{Giu98,Ber05,Bla95,Giu02B,Bla03}.
In the current experimental scenario, we can notice that the KamLAND experiment will significantly reduce the allowed region for $\Delta m^{\2}_{\1\2}$ and $\sin{(2\theta_{\1\2})}$ parameters, where the second-order wave-packet corrections can appear as an additional ingredient for accurately applying the phenomenological analysis \cite{Gro04}.
In parallel, the next major goal for the reactor neutrino program will be to attempt a measurement of $\sin^{\2}{(2\theta_{\1\3})}$, i.e. while the determination of the mixing parameters
appearing in the solar \cite{sol1,sol2} and atmospheric \cite{atm1,atm2,atm3} neutrino oscillations has already entered the precision era, the next question which can be approached experimentally is that one of $e3$ mixing.
Some important experiments which will search for more precise measurements of $\theta_{\1\3}$ are {\em Double Chooz} \cite{rea1} with designed sensitivity of $0.03$, and the {\em Daya Bay Reactor Neutrino Experiment} \cite{rea2} with designed sensitivity of $0.01$.
A consequence of a non-zero $U_{e\3}$ matrix element will be a small appearance of $\nu_{e}$ in a bean of $\nu_{\mu}$.
Assuming the scenario where $\Delta m^{\2}_{\1\2} \ll \Delta m^{\2}_{\2\3}$, which is suggested by experimental data,
and for $E_{\nu} \sim L \,\Delta m^{\2}_{\2\3}/(2\pi)$, ignoring matter effects, we find \cite{Vog04}
\small\begin{equation}
P(\mbox{\boldmath$\nu_\mu$}\rightarrow\mbox{\boldmath$\nu_e$};\,L, E_{\nu})\simeq
\sin^{\2}{(2\theta{\1\3})}\,\sin^{\2}{(\theta{\2\3})}\sin^{\2}{\left[\frac{\Delta m^{\2}_{\2\3}\, L}{4 E_{\nu}}\right]}.
\label{100}
\end{equation}\normalsize
This expression illustrates that $\theta_{\1\3}$ manifests itself in the amplitude of an oscillation
between the second and third families.
To improve the experimental limits on $\theta_{\1\3}$, one needs both good statistics and low background data.
At the same time, all kind of {\em fine-tuning} correction should also deserve a quantitative analysis.
In particular, it can be shown that reactor experiments have the potential to determine $\theta_{\1\3}$ without ambiguity from CP violation or matter effects (by assuming the necessary statistical precision which requires large reactor power and large detector size). With reasonable systematic errors ($< \,1 \%$) the sensitivity is supposed to reach $\sin^{\2}{(2\theta_{\1\3})}\approx 0.01-0.02$ \cite{Vog04} and an accurate method of analysis, maybe in the wave-packet framework, can be required.

The most simplified theoretical formulation used for describing the flavor conversion process involves the {\em intermediate} wave-packet treatment \cite{Kay81,Zra98} which eliminates the most controversial points rising up with the {\em standard} plane-wave formalism \cite{Kay81,Kay89,Kay04}\footnote{The wave-packets describing propagating mass-eigenstates guarantee the existence of a coherence length, avoid the ambiguous approximations in the plane-wave derivation of the phase difference  and, under particular conditions of minimal decoherence,
recover the flavor change probability given by the {\em standard} plane-wave treatment.}.
It is convenient to observe that the {\em intermediate} wave-packet procedure leads to flavor conversion expressions that,
after some suitable parameter adjustments,
are mathematically equivalent to those ones obtained with the average energy treatment usually applied to plane waves \cite{Kim93}.
From the point of view of a first quantized theory and in the context of vacuum oscillations, our main purpose is
to compare
the standard quantum oscillation plane wave treatment with an analytical study in the wave-packet framework by re-obtaining the
energy dependence of the oscillation probability formula
in a particular phenomenological context.
In this sense, by analytically quantifying the dependence of the neutrino oscillation parameters on the product between the wave-packet width $a$ and the average energy $\varepsilon$ of detection, we shall obtain the their range of deviation from the values obtained with the plane-wave approach.
Therefore, we suggest an improvement on bounds in adjusting the focusing horn, target position and/or detector location for some flavor conversion experiments.

In neutrino oscillation experiments, the distance of the detector from the source $L$, the neutrino average energy $E_{\nu} = \varepsilon$, and the appearance (disappearance) probability $\langle P \rangle$
are the experimental {\em input} parameters which lead to the {\em output} mixing angle and mass-difference parameters.
For discussing how the procedure for obtaining this parameters can be modified/improved,
we are effectively interested in quantifying the energy dependence of oscillation probabilities
when the {\em intermediate} wave-packet treatment is taken into account.

The first step of our study concerns the analytical derivation of a flavor oscillation formula where a
{\em gaussian} momentum distribution and a power series expansion of the energy up to the second-order terms
are utilized for obtaining analytically integrable expressions which result in the flavor conversion probabilities.
We also state that the main aspects of the oscillation phenomena can be understood by studying the two-flavor problem.
In addition, substantial
mathematical simplifications result from the assumption that the space dependence of wave functions is one-dimensional ($z$-axis).
With such simplifying hypotheses, the time evolution of flavor wave-packets can be described by
\small\begin{eqnarray}
\Phi(z,t) &=& \phi_{\1}(z,t)\cos{\theta}\,\mbox{\boldmath$\nu_{\1}$} + \phi_{\2}(z,t)\sin{\theta}\,\mbox{\boldmath$\nu_{\2}$}\nonumber\\
          &=& \left[\phi_{\1}(z,t)\cos^{\2}{\theta} + \phi_{\2}(z,t)\sin^{\2}{\theta}\right]\,\mbox{\boldmath$\nu_\alpha$}+
		  \nonumber\\& &~~~~~~~~~
		  \left[\phi_{\2}(z,t) - \phi_{\1}(z,t)\right]\cos{\theta}\sin{\theta}\,\mbox{\boldmath$\nu_\beta$}\nonumber\\
          &=& \phi_{\alpha}(z,t;\theta)\,\mbox{\boldmath$\nu_\alpha$} + \phi_{\beta}(z,t;\theta)\,\mbox{\boldmath$\nu_\beta$},
\label{0}
\end{eqnarray}\normalsize
where {\boldmath$\nu_\alpha$} and {\boldmath$\nu_\beta$} are flavor-eigenstates and {\boldmath$\nu_{\1}$} and {\boldmath$\nu_{\2}$} are mass-eigenstates.
The probability that neutrinos originally created as a $\mbox{\boldmath$\nu_\alpha$}$ flavor-eigenstate with average energy $\varepsilon$ oscillate
into a $\mbox{\boldmath$\nu_\beta$}$ flavor-eigenstate after a time $t$ is given by the
$\mbox{\boldmath$\nu_\beta$}$ coefficient
\small\begin{equation}
P(\mbox{\boldmath$\nu_\alpha$}\rightarrow\mbox{\boldmath$\nu_\beta$};t)=
\int_{_{\infm}}^{^{\infp}}\hspace{-0.5cm}dz \,\left|\phi_{\beta}\right|^{\2}
=
\frac{\sin^{\2}{(2\theta)}}{2}\left\{\, 1 - \mbox{\sc Int}(t) \, \right\},
\label{1}
\end{equation}\normalsize
where $\mbox{\sc Int}(t)$ represents the mass-eigenstate interference term given by
\small\begin{equation}
\mbox{\sc Int}(t) = Re
 \left[\, \int_{_{\infm}}^{^{\infp}}\hspace{-0.5cm}dz
\,\phi^{\dagger}_{\1}(z,t) \, \phi_{\2}(z,t) \, \right].\,
\label{2}
\end{equation}\normalsize
Since the time evolution of each mass-eigenstate wave-packet is given by \cite{Ber04B,Ber04B2}
\small\begin{eqnarray}
\phi_{\ii}(z,t) =
\int_{_{\infm}}^{^{\infp}}\hspace{-0.25cm}\frac{dp_{\z}}{2 \pi} \,
\varphi(p_{\z} \mi p_{\ii}) \exp{\left[-i\,E^{(\ii)}_{p_{\z}}\,t +i \, p_{\z}
\,z\right]},
\label{4}
\end{eqnarray}\normalsize
where
$E^{(\ii)}_{p_{\z}} = \left(p_{\z}^{\2} + m_{\ii}^{\2}\right)^{ \frac{1}{2}}$, $i = 1,\, 2$
and
$\varphi(p_{\z} \mi p_{\ii}) =  \left(2 \pi a^{\2} \right)^{ \frac{1}{4}} \exp{\left[- \frac{(p_{\z} \mi p_{\ii})^{\2}\,a^{\2}}{4}\right]}$,
we can calculate the interference term $\mbox{\sc Int}(t)$ by evaluating
the following integral
\small\begin{eqnarray}
\lefteqn{\int_{_{\infm}}^{^{\infp}}\hspace{-0.25cm}\frac{dp_z}{2 \pi} \,  \varphi(p_z \mi p_{ 1}) \varphi(p_z \mi p_{ 2})
\exp{[-i \, \Delta E_{p_z} \, t]} =}\nonumber\\
&& \exp{\left[ \frac{\mi(a \, \Delta{p})^{\2}}{8}\right]}
\int_{_{\infm}}^{^{\infp}}\hspace{-0.25cm}\frac{dp_z}{2 \pi}  \, \varphi^{\2}(p_z \mi p_{\0})\exp{[-i \, \Delta E_{p_z} \, t]},~~~ \label{6}
\end{eqnarray}\normalsize
where we have changed the $z$-integration into a $p_{\z}$-integration and introduced the quantities
$\Delta p = p_{ \1} \mi p_{ \2},\,\, p_{\0} = \frac{1}{2}(p_{ \1} + p_{ \2})$ and
$\Delta E_{p_{\z}} = E^{(\1)}_{p_{\z}} \mi E^{(\2)}_{p_{\z}}$.
The oscillation term is delimited
by the exponential function of $a \, \Delta p$ at any instant of time.
Under this condition, we
could never observe a {\em pure} flavor-eigenstate.
Besides, oscillations are considerably suppressed if $a \, \Delta p > 1$.
A necessary condition to observe oscillations is that $a \, \Delta p \ll 1$.
This constraint can also be expressed by $\delta p \gg \Delta p$ where $\delta p$ is the momentum uncertainty of the particle.
The overlap between the momentum distributions is indeed relevant only for $\delta p \gg \Delta p$.
Strictly speaking, we are assuming that the oscillation length ($\pi\frac{4 \varepsilon}{\Delta m^{\2}_{\ii\j}}$) is sufficiently larger than the wave-packet width, which
simply says that the wave-packet must not be
as wide as the oscillation length, otherwise the oscillations are washed out \cite{Kay81,Gri96,Gri99}.
Turning back to the Eq.~(\ref{6}), without loss of generality, we can assume
\small\begin{equation}
\mbox{\sc Int}(t) = Re \left\{\int_{_{\infm}}^{^{\infp}}\hspace{-0.25cm}\frac{dp_{\z}}{2
\pi}
 \, \varphi^{\2}(p_{\z} \mi p_{\0})\exp{[-i \, \Delta E_{p_{\z}} \, t]}\right\}
\label{9}.
\end{equation}\normalsize

In the literature, this equation is often obtained by assuming two mass-eigenstate wave-packets described by the same momentum distribution centered around the average momentum $\bar{p} = p_{\0}$.
This simplifying hypothesis also guarantees the
{\em instantaneous} creation of a {\em pure} flavor eigenstate {\boldmath$\nu_\alpha$} at $t = 0$ \cite{DeL04}.
In fact, we get
$\phi_{\alpha}(z,0,\theta)=\phi_{\1}(z,0)=\phi_{\2}(z,0)$  from Eq.~(\ref{0})
and
$\phi_{\beta}(z,0,\theta) =0$.
In order to obtain an expression for $\phi_{\ii}(z,t)$ by analytically evaluating
the integral in Eq.~(\ref{4}) we firstly rewrite the energy $E^{(\ii)}_{p_{\z}}$ as
$E^{(\ii)}_{p_{\z}} = E_{\ii} \left[1 + \sigma_{\ii} \left(\sigma_{\ii} + 2 \mbox{v}_{\ii}\right)\right]^{ \frac{1}{2}}$,
where $E_{\ii} = (m_{\ii}^{\2} + p_{\0}^{\2})^{ \frac{1}{2}}$, $\mbox{v}_{\ii} = \frac{ p_{\0}}{E_{\ii}}$ and
$\sigma_{\ii} = \frac{ p_{\z} \mi  p_{\0}}{E_{\ii}}$.
The integral in Eq.~(\ref{4}) can be {\em
analytically} evaluated only if we consider terms up to order
$\sigma_{\ii}^2$ in a power series expansion conveniently truncated as
\small\begin{eqnarray}
E^{(\ii)}_{p_{\z}} & = & E_{\ii} \left[1 + \sigma_{\ii} \mbox{v}_{\ii}  + \frac{\sigma_{\ii}^{\2}}{2}\left(1 - \mbox{v}_{\ii}^{\2} \right)\right] + \mathcal{O}(\sigma_{\ii}^{\3})
\nonumber\\  &
 = &
  E_{\ii} +  p_{\0} \sigma_{\ii} + \frac{m_{\ii}^{\2}}{2E_{\ii}} \sigma_{\ii}^{\2} + \mathcal{O}(\sigma_{\ii}^{\3}).
\label{12}
\end{eqnarray}\normalsize
The zero-order term $E_{\ii}$ in the above expansion gives the standard plane-wave oscillation phase.
The first-order term $p_{\0}\sigma_{\ii}$ is responsible for the {\em slippage} between the mass-eigenstate wave-packets due to their different group velocities.
It represents a linear correction to the standard oscillation phase.
Finally, the second-order term $\frac{m_{\ii}^{\2}}{2E_{\ii}} \sigma_{\ii}^{\2}$, which is a (quadratic) secondary correction,
gives the well known {\em spreading} effects in the time propagation of the wave-packet.
Moreover, it leads to the appearance of an {\em additional} time-dependent phase in the final expression for the oscillation probability.
In case of {\em gaussian} momentum distributions, all these terms can be {\em analytically} quantified.
By evaluating the integral (\ref{9}) with the approximation (\ref{12}), and after
performing some mathematical manipulations \cite{Ber06,Ber05B}, we can express the interference term as
\small\begin{equation}
\mbox{\sc Int}(t) = \mbox{\sc Dmp}(t) \times \mbox{\sc Osc}(t),
\label{20}
\end{equation}\normalsize
which was factorized into a decoherence damping term
given by
\small\begin{equation}
\mbox{\sc Dmp}(t) = \left[1 + \mbox{\sc Sp}^{\2}(t) \right]^{-\frac{1}{4}}
\exp{\left[-\frac{(\Delta \mbox{v} \, t)^{\2}}{2a^{\2}\left[1 + \mbox{\sc Sp}^{\2}(t)\right]}\right]}
\label{21}
\end{equation}\normalsize
and a time-oscillating flavor conversion term given by
\small\begin{eqnarray}
\mbox{\sc Osc}(t) &=& Re \left\{\exp{\left[-i\Delta E \, t -i \Theta(t)\right]} \right\}\nonumber\\
&=& \cos{\left[\Delta E \, t + \Theta(t)\right]},
\label{22A}
\end{eqnarray}\normalsize
where
\small\begin{equation}
\mbox{\sc Sp}(t) = \frac{t}{a^{\2}}\Delta\left(\frac{m^{\2}}{E^{\3}}\right) = \rho\, \frac{\Delta \mbox{v}\, t}{a^{\2} \,  p_{\0}}
\label{240}
\end{equation}\normalsize
and
\small\begin{equation}
\Theta(t) = \mbox{$\left[\frac{1}{2}\arctan{\left[\mbox{\sc Sp}(t)\right]} - \frac{a^{\2} \,  p_{\0}^{\2}}{2 \rho^{\2}}\frac{\mbox{\sc Sp}^{\3}(t)}{\left[1 + \mbox{\sc Sp}^{\2}(t)\right]}\right]$},
\label{24A}
\end{equation}\normalsize
with
$\rho = 1 - \left[3 + \left(\frac{\Delta E}{\varepsilon}\right)^{\2}\right] \frac{ p_{\0}^{\2}}{\varepsilon^{\2}}$,
$\Delta E = E_{\1} - E_{\2}$ and $\varepsilon = \sqrt{E_{\1} \, E_{\2}}$.
The time-dependent quantities $\mbox{\sc Sp}(t)$ and $\Theta(t)$
carry the second-order corrections and, consequently, the
{\em spreading} effect to the oscillation probability formula \cite{Ber04B,Ber04B2}.
If $\Delta E \ll \varepsilon$, the parameter $\rho$ is limited by
the interval $[1,-2]$ and it assumes the zero value when $\frac{ p_{\0}^{\2}}{\varepsilon^{\2}} \approx \frac{1}{3}$.
The {\em slippage} between the mass-eigenstate wave-packets is
quantified by the vanishing behavior of the damping term $\mbox{\sc Dmp}(t)$.
The NR limit is obtained by setting $\rho^{\2} = 1$ and $ p_{\0} = 0$ in Eq.~(\ref{21}).
In the same way, the UR limit is obtained by setting $\rho^{\2} = 4$
and $ p_{\0} = \varepsilon$.
In fact, the minimal influence due to second-order corrections occurs
when $\frac{ p_{\0}^{\2}}{\varepsilon^{\2}} \approx \frac{1}{3}$ ($\rho \approx 0$).
Returning to the exponential term of Eq.~(\ref{21}), we observe that the oscillation amplitude is more
relevant when $\Delta \mbox{v} \, t \ll a$.
It characterizes the {\em minimal slippage} between the mass-eigenstate
wave-packets which occur when the
complete spatial intersection between themselves starts to diminish
during the time evolution.

The oscillating component $\mbox{\sc Osc}(t)$ of the interference
term $\mbox{\sc Int}(t)$ differs from the {\em standard} oscillating
term $ \cos{[\Delta E \, t]}$
by the presence of the residual phase $\Theta(t)$,
which is essentially a second-order correction \cite{Ber04B,Ber04B2}.
Superposing the effects of $\mbox{\sc Dmp}(t)$ and the oscillating character $\mbox{\sc Osc}(t)$,
we immediately obtain the flavor oscillation probability in its explicit form \cite{Ber06,Ber05B},
\small\begin{eqnarray}
\lefteqn{
P(\mbox{\boldmath$\nu_\alpha$}\rightarrow\mbox{\boldmath$\nu_\beta$};t)
 \approx
\frac{\sin^{\2}{(2\theta)}}{2}\left\{1 - \left[1 + \mbox{\sc Sp}^{\2}(t) \right]^{-\frac{1}{4}}
\right.
}\nonumber\\&~~~~~~~~~\times
\left.
\exp{\left[-\frac{(\Delta \mbox{v} \, t)^{\2}}{2a^{\2}\left[1 + \mbox{\sc Sp}^{\2}(t)\right]}\right]}
\cos{\left[\Delta E \, t + \Theta(t)\right]}
  \right\},
\label{25A}
\end{eqnarray}\normalsize
from which we notice that the larger is the value of $a \varepsilon$, the smaller are the wave-packet effects.

To perform some phenomenological analysis involving the $\theta_{\1\3}$ mixing angle,
we replace $\sin^{\2}{(2\theta)}$ by $\sin^{\2}{(2\theta_{\1\3})}\sin^{\2}{(\theta_{\2\3})}$ and
we set $\Delta E \equiv \Delta E_{\2\3}$ in the Eq.~(\ref{25A}) in order to realistically characterize
the $\nu_{\mu}\rightarrow\nu_{e}$ conversion which emerges in a three flavor scenario.
We establish the experimental {\em input} parameters as being the distance $L$, the neutrino energy
distribution $\varepsilon$ and the appearance (disappearance) probability $\langle P \rangle$.
At this point, it is instructive to redefine
some parameters which shall carry the main physical information in the oscillation formula.
Firstly, we set the oscillation length {\em scale} $L_{\0}$ which is related to an energy {\em scale} $E_{\0}$
by the expression $L_{\0} = 2 \pi \frac{E_{\0}}{\Delta m^{\2}_{\2\3}}$.
Both parameters, $L_{\0}$ and $E_{\0}$, correspond to referential scales that can be calibrated in agreement with the experimental configuration and the data analyzed.
From a practical point of view, the criteria for the choice of this parameters is not so arbitrary.
In a phenomenological analysis, the choice of the parameter $E_{\0}$ can be done in correspondence with the
peak of an energy distribution ($\varepsilon$) of a certain type of neutrino flux for which the experimentally obtained energy distribution is typically determined by the neutrino production processes.
As we shall observe in the analysis which follows the calculations, in order to quantify the corrections due to the wave packet approximation,
the reference value of $E_{\0}$ can also be set equal to an averaged value $\langle \varepsilon \rangle \equiv \bar{E}$ where,
depending on the width of the energy distribution of the neutrino flux,
the necessity of an additional energy integration over $\varepsilon$ (averaged out integration) is discarded.

We also introduce the auxiliary definitions $\delta = a \varepsilon$ and $\upsilon = \frac{p_{\0}}{\varepsilon}$
which respectively parameterize the wave-packet character and the propagation regime.
With the previous definition of $\varepsilon$,
we introduce the dimensionless variables,
\small\begin{equation}
x = \frac{\varepsilon}{E_{\0}}~~~~ \mbox{and}~~~~l = \frac{L}{L_{\0}}
\label{pap1A}
\end{equation}\normalsize
which will be useful in the the subsequent analysis, since it allows us to extend the validity of the interpreted results to any set of parameters $L_{\0}$ and $E_{\0}$.
In real experiments the neutrino energy, $\varepsilon$, and sometimes the detection position, $\upsilon t \, \sim L$, can have some spread around and/or deviation from
respectively $E_{\0}$ and $L_{\0}$ due to various effects, but in a subset
of this experiments there is a well-defined value of $\langle L/\varepsilon \rangle\sim L_{\0}/E_{\0}$
(or $x/l \sim 1$ in the plane-wave limit as we shall see in the following) around which the events distribute.
Following the same approach that
we have adopted while we were analyzing
the parameter $\rho$ in Eq.~(\ref{240}), if  $\Delta E \ll \varepsilon$, which is perfectly acceptable from the
experimental point of view, we can write $\varepsilon = \sqrt{E_{\1} {E_{\2}}}
\approx \frac{1}{2}(E_{\1}+{E_{\2}})$ so that an effective plane-wave flavor conversion
formula can be obtained from Eq.~(\ref{100}) as
\small\begin{eqnarray}
\langle P \rangle_{\mbox{\tiny PW}} & \equiv & P(\mbox{\boldmath$\nu_\mu$}\rightarrow\mbox{\boldmath$\nu_e$};l) \nonumber\\
& = & \sin^{\2}{(2\theta{\1\3})}\,\sin^{\2}{(\theta{\2\3})} \left\{1 - \cos{\left[\frac{\pi\, l}{x}\right]}\right\}
\label{pap2}.
\end{eqnarray}\normalsize
Analogously, we can observe by means of the Eqs~(\ref{20}-\ref{24A}) that the wave-packet flavor conversion formula with second-order corrections (\ref{25A})
exhibits a similar implicit dependence on time.
The Eq.~(\ref{25A}) can thus be rewritten as a function of the above parameter $x$ (\ref{pap1A}) in terms of
\small\begin{equation}
\mbox{\sc Dmp}(x) = \left[1 + \mbox{\sc Sp}^{\2}(x) \right]^{\mi\frac{1}{4}}\exp{\left[- \left(\frac{\pi\, l}{2\,\delta\,x\,\upsilon}\right)^{\2}\frac{2}{1 + \mbox{\sc Sp}^{\2}(x)}\right]},
~~\label{pap3}
\end{equation}\normalsize
and
\small\begin{eqnarray}
\mbox{\sc Osc}(x) &=& \cos{\left[\frac{\pi\, l}{x} + \Theta(x)\right]},
\label{pap4}
\end{eqnarray}\normalsize
where
\small\begin{equation}
\mbox{\sc Sp}(x) = - \rho \frac{\pi\, l}{x\, \delta^{\2}}
\label{pap5}
\end{equation}\normalsize
and
\small\begin{equation}
\Theta(x) = \left[\frac{1}{2}\arctan{\left[\mbox{\sc Sp}(x)\right]}
- \frac{\delta^{\2} \,  \upsilon^{\2}}{2 \rho^{\2}}
\frac{\mbox{\sc Sp}^{\3}(x)}{\left[1 + \mbox{\sc Sp}^{\2}(x)\right]}\right],
\label{pap6}
\end{equation}\normalsize
with $\rho \approx 1 - 3 \upsilon^{\2}$.
We can observe that the parameter $\delta =  a \varepsilon$ carries
the relevant information about the wave-packet width $a$ and the average energy $\varepsilon$.
If it was sufficiently large ($\delta \gg 1$) so that we could ignore
the second-order corrections in Eq.~(\ref{12}), the probability with the leading terms
could be read as
\small\begin{eqnarray}
\lefteqn{
\langle P \rangle_{\mbox{\tiny WP1}} = \sin^{\2}{(2\theta{\1\3})}\,\sin^{\2}{(\theta{\2\3})}
}\nonumber\\&&~~~\times
\left\{1-\cos{\left[\frac{\pi\, l}{x}\right]}\,
\exp{\left[- 2 \left(\frac{\pi \,l}{2\,\delta\,x\,\upsilon}\right)^{\2}\right]}\right\},\label{pap7}
\end{eqnarray}\normalsize
which, in the particular case of an ultra-relativistic propagation ($\upsilon = 1$),
can be used as a reference for comparison with experimental data \cite{Gro04}.
By the way, despite the relevant dependence on the propagation regime ($\upsilon$),
once we are interested in some realistic physical situations,
the following analysis will be limited to the ultra-relativistic propagation regime corresponding to
the effective neutrino energy of the current flavor oscillation experiments.

\begin{figure}
\begin{center}
\hspace{-0.7cm}
\epsfig{file= 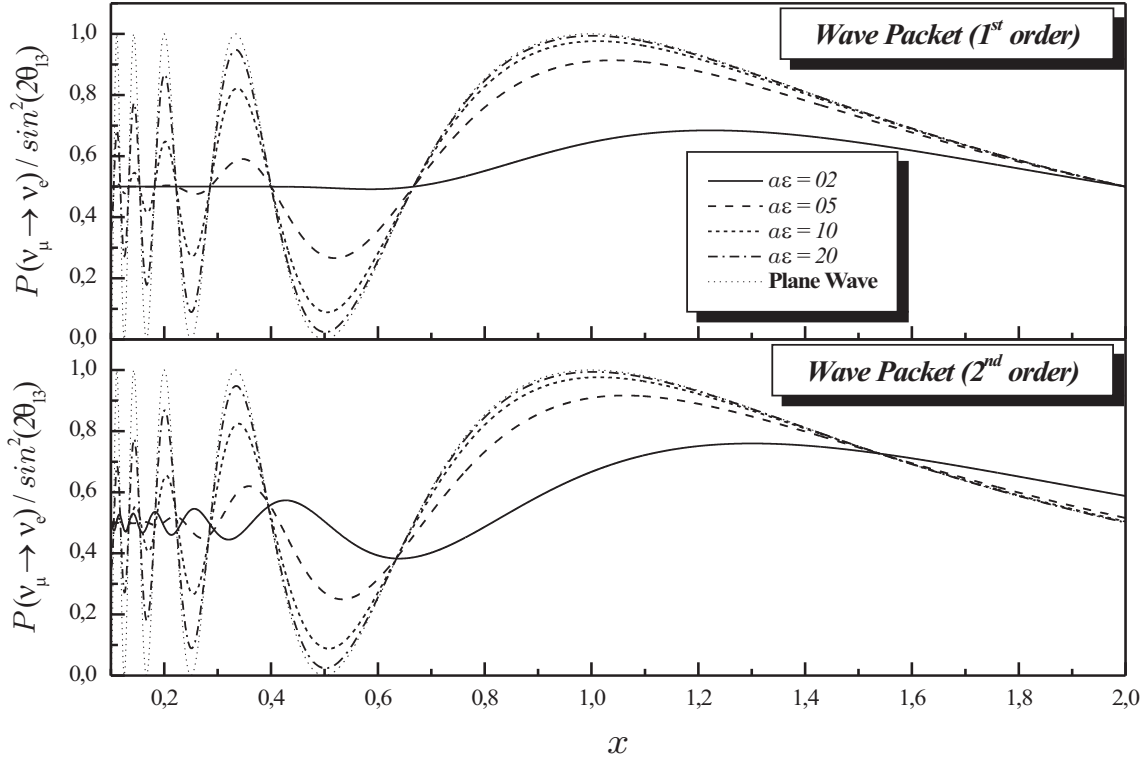, height= 12 cm, width= 17 cm}
\end{center}
\vspace{-1 cm}
\caption{\label{Fig1}
Fixed-distance probabilities $P(\mbox{\boldmath$\nu_\mu$}\rightarrow\mbox{\boldmath$\nu_e$})$ at $l = \frac{L}{L_{\0}} = 1$, normalized by $\sin^{\2}{(2\theta_{\1\3})}$, with $\theta_{\2\3} = \pi/4$,
as a function of the dimensionless energy $x = \frac{\varepsilon}{E_{\0}}$.
We have assumed that the wave-packet approximation is fixed by
$\delta = a \varepsilon = 2,\,5,\,10,\, 20$ for any arbitrary value of $x(\varepsilon)$.
For sufficiently large values of $\delta$, for instance when $\delta = a \varepsilon = 20$, we recover the plane wave result which, at first glance (visually),
coincides with the $1^{st}$ (first plot) $2^{nd}$ (second plot) order results respectively given by Eqs.~(\ref{pap7}) and (\ref{25A}).
Just for completeness, the same map can be reproduced when we set $x \rightarrow x/l$ for unconstrained $l$,
i.e. instead of assuming $l = 1$ which sets the values of $L_{\0}$ and $E_{\0}$ separately, we
can choose to fix the ratio $\langle L_{\0}/E_{\0}\rangle$ eliminating one degree of freedom.
It allows us to extend the validity of the information that we can extract from the figure to
a larger set of parameters $L_{\0}$ and $E_{\0}$ which characterizes an experimental apparatus.}
\end{figure}

\begin{figure}[h]
\begin{center}
\hspace{-0.7cm}
\epsfig{file= 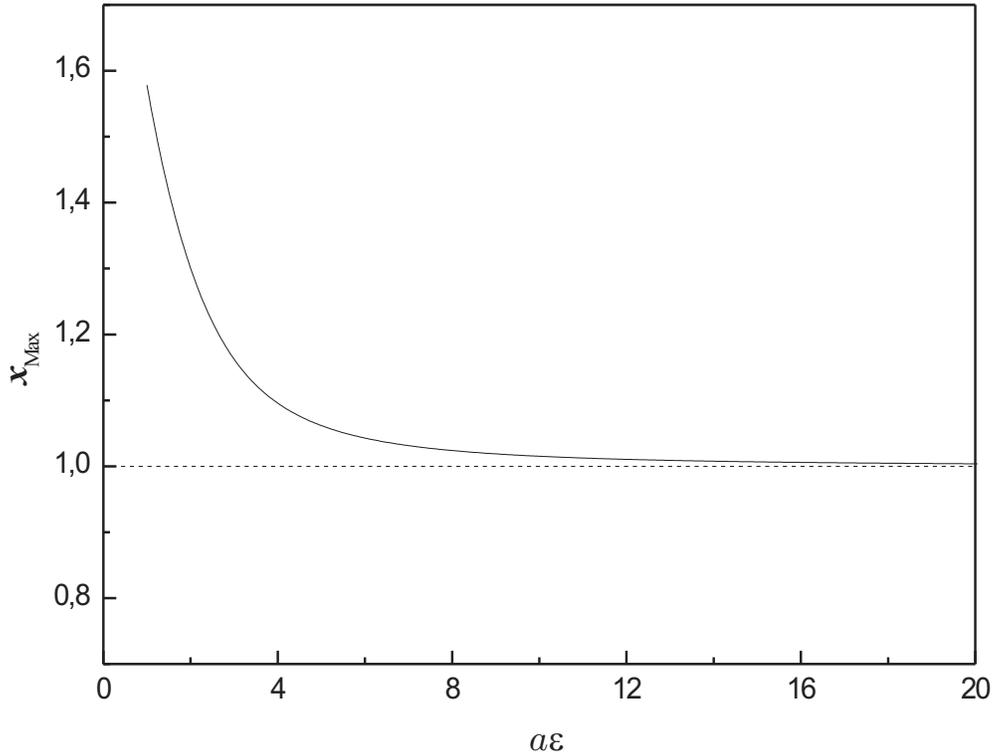, height= 12 cm, width= 15 cm}
\end{center}
\vspace{-1 cm}
\caption{\label{Fig2}
The first maximum $x_{\mbox{\tiny Max}}$ of the probability expression $P(\mbox{\boldmath$\nu_\mu$}\rightarrow\mbox{\boldmath$\nu_e$})$ as a function of the wave-packet
correction  parameter $a \varepsilon$ varying from $1$ to $20$.
For plane waves the maximum occurs at $x_{\mbox{\tiny Max}} = 1$.}
\end{figure}

\begin{figure}
\begin{center}
\hspace{-0.5cm}
\epsfig{file= 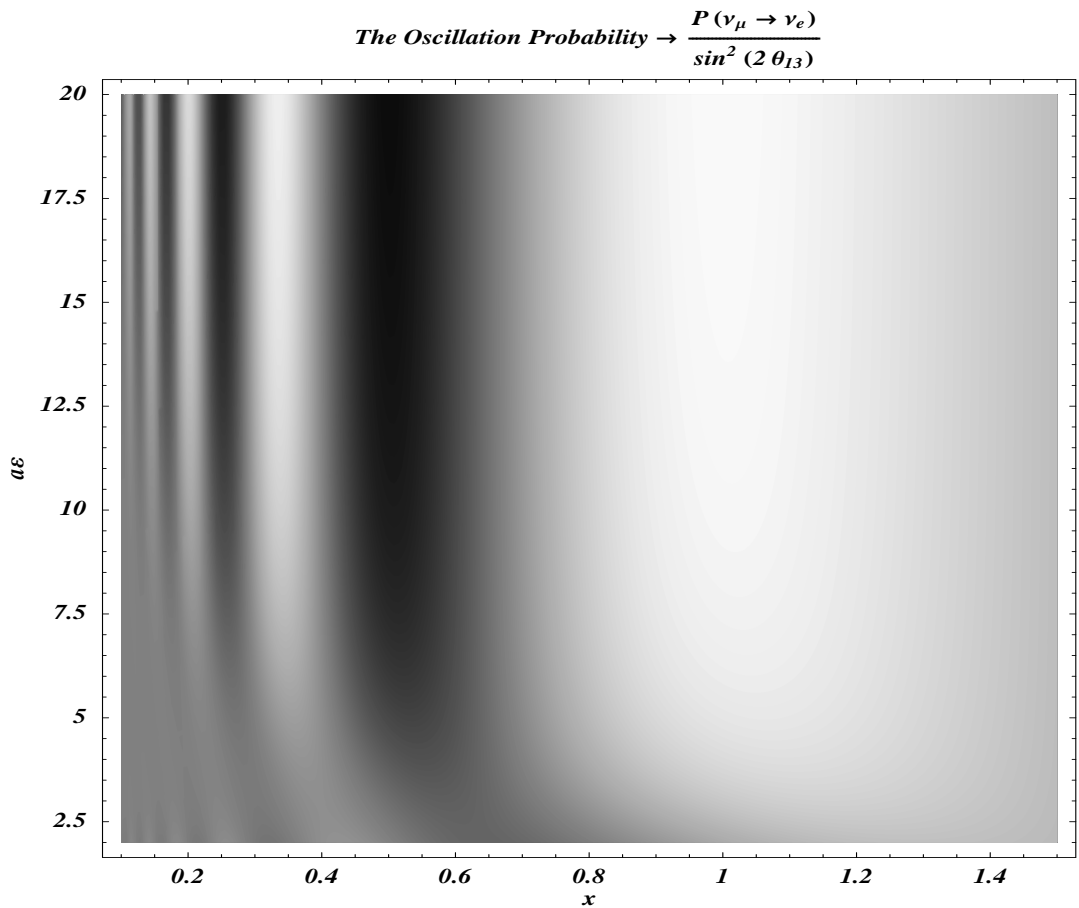, height= 12 cm, width= 15 cm}
\end{center}
\vspace{-0.6cm}
\caption{\label{Fig4}
The contour plot corresponding to the 3-dimensional representation of the dependence of fixed-distance probabilities $P(\mbox{\boldmath$\nu_\mu$}\rightarrow\mbox{\boldmath$\nu_e$})$
(at $l = \frac{L}{L_{\0}}= 1$) on the parameter $a \varepsilon$ which characterizes the dependence on the wave-packet approximation.
This plot illustrates the magnitude of the oscillations in a map of $x(\varepsilon)$ by $a \varepsilon$.
Darker corresponds to small probabilities.}
\end{figure}

As previously mentioned, the shape of the oscillation probability curve as a function of the energy ($x$)
for the above approximations is, indeed, different from that one
of the standard plane-wave treatment,
as we can observe in the Fig.~(\ref{Fig1})
where we have plotted the fixed-distance probabilities
$P(\mbox{\boldmath$\nu_\mu$}\rightarrow\mbox{\boldmath$\nu_e$})$ normalized by $\sin^{\2}{(2\theta_{\1\3})}$
as a function of the dimensionless energy $x = \frac{\varepsilon}{E_{\0}}$ for four different values of $\delta = a \varepsilon$.
In the first plot we illustrate the wave-packet approximation with $1^{st}$ order corrections
parameterized by the Eq.~(\ref{pap7}) and
in the second plot we illustrate to the wave-packet approximation with $2^{nd}$ order corrections
parameterized by the Eq.~(\ref{25A}) where the dependence on $x$ is implicit.
The quoted approximations can be compared with the plane-wave approximation (dotted line).
In order to keep clear the meaning of the deviation of the wave-packet approximation from the plane-wave approximation,
in spite of the dependence on the energy of the parameter $\delta =  a \varepsilon$,
we are constrained to set constant values to it for each curve which expresses
the probability dependence on the energy.
Alternatively, we could set $\delta_{\0} = a \, E_{\0}$ and $ \delta = x \, \delta_{\0}$ in order to re-plot
the oscillation probability dependence on $x$, which is, however, completely unrealistic under the point of view of the approximation accurateness.
The correction on the first maximum of probability that allows us
to adjust the focusing horn, target position and/or detector location for some flavor conversion experiments
is represented in the Fig.~\ref{Fig2} where the maximal values of $x$ were
numerically obtained as a function of $a \varepsilon$.
Considering the energy dependence represented in the Fig.~(\ref{Fig1}), it is advantageous to introduce
a third axis representing the dependence on the parameter
$a \varepsilon$ in order to illustrate the complete/effective oscillation behavior.
The Fig.~(\ref{Fig4}) allows us to qualitatively identify the influence of the wave-packet
corrections brought up by $a \varepsilon$.
Quite generally, the complete analysis of the oscillating character coupled to the loss of coherence between the mass-eigenstate wave-packets, which suppresses the flavor oscillation amplitude, depends on the experimental features such as the size of the source, which allows estimating the wave-packet width ($a$), the neutrino energy distribution ($\varepsilon$), and the detector resolution ($L_{\0}$).
Once they produce an effect competing with that of the finite size of the wave-packet, the neutrino energy measurements cannot be performed very precisely.
If we set the energy uncertainty represented by $\delta E$, the Heisenberg uncertainty relation states that $\delta E\, a \sim 1$ and, consequently, our approximation hypothesis leads to $\frac{\delta E}{\bar{E}} \sim \frac{1}{a\,\bar{E}} \ll 1$.
Realistically speaking, a typical neutrino-oscillation experiment searches for flavor conversions by means of
an apparatus which, apart from the details inherent to the physical process, provides an indirect measurement of the neutrino energy (in each event) accompanied by an experimental error $\Delta E_{exp}$ due to the ``detector resolution''.
In case of $\frac{\Delta E_{exp}}{\bar{E}}<\frac{\delta E}{\bar{E}} \sim \frac{1}{a\,\bar{E}}$, the effective role of the second-order corrections illustrated in this analysis can be relevant
since, as we have anticipated, the necessity of an additional energy integration over the energy distribution $\varepsilon$ (averaged out integration) is discarded.
On the contrary, $\Delta E_{exp} > \delta E$ demands for an average energy integration where the decoherence effect through imperfect neutrino energy measurements by far dominates.
In this sense, the current experimental values/measurements set some limitations on the applicability of our analysis which, at this point, is restricted to the $^{\7}Be$ and $pep$ lines for solar neutrinos ($\frac{\Delta E_{exp}}{\bar{E}} \ll 1$), certainly to some (next generation) reactor experiment where the designed sensitivity is of the order of $(a \varepsilon)^{\mi\1}$, and eventually to supernova neutrinos \footnote{More discussion about the choice of $a \varepsilon$ is done in \cite{Ber06}.}.

Generically speaking, although the higher energy neutrinos are more accessible experimentally, the corrections to the wave packet formalism can be physically relevant for $p-p$ neutrinos with energy distributed around the values of $\varepsilon \approx 10-100 \,keV$.
Following the standard procedure \cite{Kim93} (which is not free of controversial criticisms) for calculating the wave packet width $a$ of the neutrino flux for $p-p$ solar reactions, we obtain $a = 10^{\mi 10}- 10^{\mi 8}\, m \equiv 0.5-50\, (keV)^{\mi 1}$.
Such an interval sets a very particular range for the $\sigma$ parameter comprised by the interval $5\cdot 10^{\mi \5} - 0.2$ for $\varepsilon \approx 0.01-0.4\, MeV$ which introduces the possibility for wave packet second-order corrections establish some not ignoble modifications for the $p-p$ neutrino oscillation parameter limits.
In a supernova, the size of the wave packet is determined by the region of production (plasma), due to a process known as pressure broadening, which depends on the temperature, the plasma density, the mean free path of the neutrino producing particle and its mean termal velocity \cite{Kim93}.
Neutrinos from supernova core with $100\,MeV$ energy have a wave packet size varying from $\sim 5 \cdot 10^{\mi \1 \6} \,m$ to $\sim 10^{\mi \1 \4} \,m$ which leads to a wave packet parameter $a \varepsilon$ comprised by the interval $0.25 - 5$ for which the second-order corrections can be indeed relevant.
In fact, once we have precise values for the input parameters $\langle P \rangle$, $L$ and $\varepsilon$, we could determine the effectiveness of the first/second-order corrections in determining $\Delta m^{2}$ for any class of neutrino oscillation experiment.
For instance, the flux of atmospheric neutrinos produced by collisions of cosmic rays (which are mostly protons) with the upper atmosphere is measured by experiments prepared for observing $\nu_{e} \leftrightarrow \nu_{\mu}$ and $\bar{\nu}_{e} \leftrightarrow \bar{\nu}_{\mu}$ conversions.
The neutrino energies range about from $0.1\, GeV$ to $100\, GeV$ which constrains the relevance of WP effects to an wave packet width $a \sim 10^{\mi\1\2}\, m$.
The majority of the old generation of the reactor neutrino experiments, cover a large variety of neutrino flavor conversions where the neutrino energy flux times the corresponding wave packet width $a$ makes the wave packet second-order corrections, at first glance, not so relevant ($a \varepsilon >> 1$ tends to the plane wave limit).

As an additional remark, it is pertinent to emphasize that there is no accurate way to experimentally measure or phenomenologically compute
the wave-packet width of a certain type of neutrino flux, for which we have only crude estimations.
Consequently, apart from the {\em obvious} criticisms to the plane-wave approach \cite{Kay81},
we cannot arbitrarily assume that the modifications introduced by the wave-packet treatment (in particular, with second-order corrections)
are irrelevant in the analysis of any generic class of neutrino experimental data.
Maybe, in a very particular scenario, the above analysis can be applied in designs of
some experiment dedicated to the $\theta_{\1\3}$ mixing angle measurement.
Finally, from the phenomenological point of view, the general arguments presented in \cite{Gro04} continue to be valid, i.e. the above discussion has so far been limited to {\em vacuum} oscillations.
In conclusion, the characterization of the wave-packet
($a$) (implicitly described by $\sigma$) accompanied by the precise determination of the neutrino energy
distribution ($\varepsilon$) should be considered when the {\em accuracy} in obtaining the neutrino oscillation parameters or their limits is the subject of the phenomenological analysis.

\begin{acknowledgments}
This work was partially supported by FAPESP (PD 04/13770-0) and CNPq.
\end{acknowledgments}


\begin{thebibliography}{99}
\bibitem{sol1}
B. Aharmim {\it et al.} [SNO Collaboration], Phys. Rev. {\bf C72}, 055502 (2005);
\bibitem{sol2}
Q. R. Ahmad {\it et al.} [SNO Collaboration], Phys. Rev. Lett {\bf 89},  011302 (2002); {\em ibid.}, Phys. Rev. Lett {\bf 89},  011301 (2002); {\em ibid.}, Phys. Rev. Lett {\bf 87}, 071301 (2001).
\bibitem{atm1}
J. Hosaka {\it et al.} [Superkamiokande Collaboration], Phys. Rev. {\bf D74}, 032002 (2006);
\bibitem{atm2}
Y. Ashie {\it et al.} [Superkamiokande Collaboration], Phys. Rev. Lett. {\bf 93} 101801 (2004); {\em ibid.}, Phys. Rev. {\bf D71}, 112005 (2005)
\bibitem{atm3}
S. Fukuda {\it et al.} [Superkamiokande Collaboration], Phys. Lett. {\bf B539}, 179, (2002); {\em ibid.}, Phys. Rev. Lett. {\bf 86}, 5651, (2001), {\em ibid.} Phys. Rev. Lett. {\bf 85}, 3999, (2000).
\bibitem{Boe01}
F. Boehm {\it et al.}, {\em ``Palo Verde Reactor Experiment''}, Phys. Rev. {\bf D64}, 112001 (2001).
\bibitem{Egu031}
T. Araki {\it et al.} [KamLAND Collaboration], Phys. Rev. Lett. {\bf 90}, 081801 (2005);
\bibitem{Egu032}
K. Egushi {\it et al.} [KamLAND Collaboration], Phys. Rev. Lett. {\bf 90}, 021802 (2003).
\bibitem{Mes06}
M. D. Messier, {\em Review of Neutrino Oscillations Experiments}, arXiv:hep-ex/0606013, and references therein.
\bibitem{rea1}
F. Ardellier {\it et al.} [Double Chooz collaboration], {\em Letter of Intent for Double-CHOOZ: a Search for the Mixing Angle Theta13}, arXiv:hep-ex/0405032.
\bibitem{rea2}
S. Kettel {\it et al.} [Daya Bay collaboration], {\em A Precision Measurement of the Neutrino Mixing Angle theta13 using Reactor Antineutrinos at Daya Bay}, arXiv:hep-ex/0701029.
\bibitem{Zub98}
K. Zuber, Phys. Rep. {\bf 305}, 295 (1998).
\bibitem{Alb03}
W. M. Alberico and S. M. Bilenky, Prog. Part. Nucl. {\bf 35}, 297 (2004).
\bibitem{Vog04}
R. D. McKeown and P. Vogel, Phys. Rep. {\bf 395}, 315 (2004).
\bibitem{Beu03}
M. Beuthe, Phys. Rep. {\bf 375}, 105 (2003).
\bibitem{Giu98}
C. Giunti and C. W. Kim, Phys. Rev. {\bf D58}, 017301 (1998).
\bibitem{Ber05}
A. E. Bernardini and S. De Leo, Phys. Rev. {\bf D71}, 076008 (2005).
\bibitem{Bla95}
M. Blasone and G. Vitiello, Ann. Phys. {\bf 244}, 283 (1995).
\bibitem{Giu02B}
C. Giunti, {\em JHEP} {\bf 0211}, 017 (2002).
\bibitem{Bla03}
M. Blasone, P. P. Pacheco and H. W. Tseung, Phys. Rev. {\bf D67}, 073011 (2003).
\bibitem{Gro04}
D. Gromm, p.451 in Particle Data Group, S. Eidelman {\it et al.},
Phys. Lett. {\bf B592}, 1 (2004).
\bibitem{Kay81}
B. Kayser,  Phys. Rev. {\bf D24}, 110 (1981).
\bibitem{Zra98}
M. Zralek, Acta Phys. Polon.{\bf B29}, 3925 (1998)
\bibitem{Kay89}
B. Kayser, F. Gibrat-Debu and F. Perrier, {\em The Physics of Massive Neutrinos} (Cambridge University Press, Cambridge, 1989).
\bibitem{Kay04}
B. Kayser, p.145 in Particle Data Group, S. Eidelman {\it et al.}, Phys. Lett. {\bf B592}, 1 (2004).
\bibitem{Kim93}
C. W. Kim and A. Pevsner, {\em Neutrinos in Physics and Astrophysics}, (Harwood Academic Publishers, Chur, 1993).
\bibitem{Ber06}
A. E. Bernardini, M. M. Guzzo and F. R. Torres, Eur. Phys. J {\bf C48}, 613 (2006).
\bibitem{Ber05B}
A. E. Bernardini, EuroPhys. Lett. {\bf 73}, 157 (2006).
\bibitem{Ber04B}
A. E. Bernardini and S. De Leo, Eur. Phys. J. {\bf C37}, 471 (2004).
\bibitem{Ber04B2}
A. E. Bernardini and S. De Leo, Phys. Rev. {\bf D70}, 053010 (2004).
\bibitem{Gri96}
W. Grimus and P. Stockinger, Phys. Rev. {\bf D54}, 3414 (1996).
\bibitem{Gri99}
W. Grimus, P. Stockinger and S.Mohanty, Phys. Rev. {\bf D59}, 013011 (1999).
\bibitem{DeL04}
S. De Leo, C. C. Nishi and P. Rotelli, Int. J. Mod. Phys. {\bf A19}, 677 (2004).

\end{thebibliography}
\end{document}